# Case ID detection based on time series data
# - the mining use case

Edyta Brzychczy, Tomasz Pełech-Pilichowski, Ziemowit Dworakowski

AGH University of Krakow al. A. Mickiewicza 30,
30-059 Krakow, Poland
{brzych3,tomek,zdw}@agh.edu.pl

**Abstract.** Process mining gains increasing popularity in business process analysis, also in heavy industry. It requires a specific data format called an event log, with the basic structure including a case identifier (case ID), activity (event) name, and timestamp. In the case of industrial processes, data is very often provided by a monitoring system as time series of low level sensor readings. This data cannot be directly used for process mining since there is no explicit marking of activities in the event log, and sometimes, case ID is not provided. We propose a novel rule-based algorithm for identification patterns, based on the identification of significant changes in short-term mean values of selected variable to detect case ID. We present our solution on the mining use case. We compare computed results (identified patterns) with expert labels of the same dataset. Experiments show that the developed algorithm in the most of the cases correctly detects IDs in datasets with and without outliers reaching F1 score values: 96.8% and 97% respectively. We also evaluate our algorithm on dataset from manufacturing domain reaching value 92.6% for F1 score.

## 1. Introduction

Industry 4.0 solutions like autonomous robots, the Internet of Things, or cloud computing bring new process analysis and improvement opportunities. These opportunities are related to the vast amount of various sensor data generated by machinery and devices involved in industrial process execution, which contains knowledge about process performance [1].

Using raw sensor data in process analysis is one of the most recent challenges in the Business Process Management (BPM) domain [2]. This challenge is related to the existing gap between the low grain level of sensor data and the high abstraction level of a process model [3]. Therefore, sensor data requires translation before its usage to process analytics on a higher level.

The most up-to-date process analytics nowadays is process mining (PM). PM enables automated process model discovery, conformance checking, as well as process performance analysis based on event data. The event data for PM are typically structured in the event log format, which in its basic form contains the following attributes: *case ID*, *activity* (event) name, and *timestamp*. The presence of a case ID is one of the essential requirements for an event log, next to the activity name executed within the process [4].

Considering the sensor data, these two essential event log elements are often not straightforwardly available in raw data. Thus, they need to be defined based on supervised or unsupervised techniques, taking into consideration the specificity of the process domain.

Our work focuses on case ID detection, known as the event correlation task [5]. Event correlation aims at associating event data extracted from data sources to cases of a business process [6]. This task is often subjective and primarily affected by domain knowledge [7].

There are many approaches to event correlation described in literature (discussed in more detail in Section 3); however, most of them assume that defined events exist in data. Unfortunately, in the case of raw industrial sensor data, only basic sensor readings are available, making case ID detection a challenging and complex task.

Our main contribution in this paper is a development of a case ID detection algorithm based on time-series patterns in sensor data. Our work was motivated by an industrial use case from the mining domain, namely the longwall shearer operation process, but the proposed method can be extended to all cases when process is running in a cyclic manner without clearly noticed start and end of the case.

The paper is structured as follows. Section 2 describes our use case and hitherto basic heuristic for case ID detection. Section 3 presents related work to the event correlation task. Section 4 contains details of our algorithm. Section 5 presents the results of the experiments. Section 6 contains evaluation of our approach. Section 7 concludes the paper and highlights future work.

## 2. The mining use case

Longwall mining is one of the most popular mining methods in hard coal mines. In longwall mining the coal cutting occurs at the longwall face, where the mining process is executed by a particular set of machinery, including a longwall shearer, an armoured face conveyor, mechanised roof supports, and a chain conveyor with a crusher (beam stage loader). The main process execution relies on longwall shearer operations; it moves from the beginning of the longwall face to the end of the longwall face in a cycle consisting of 16 specific technological stages (Fig.1):

- Cutting operations at the beginning of the longwall (marked as 1,5,15), in the middle of the longwall (6,14), and at the end of the longwall (7,9,13),
- Returns to the drives (3,11),
- Stoppages at the beginning of the longwall (2,4,16) and at the end of the longwall (8,10,12).

From the mining process analysis point of view, the presented shearer cycle is a natural candidate as an identifier of process execution traces (shearer cycles are sequential).

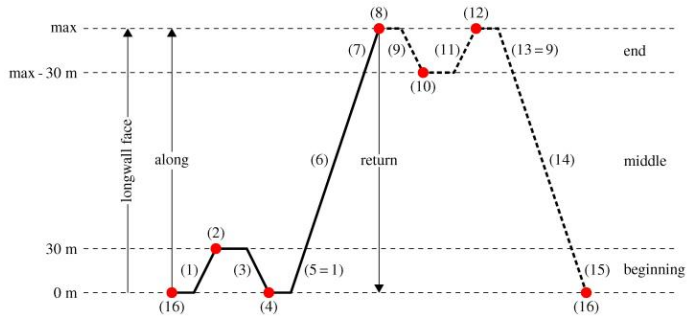

**Fig.1**: Longwall shearer cycle. Source [8]

The operations of the shearer are monitored by sensors installed on machinery and in the excavation. The main attributes collected by the monitoring system include, among others, currents on drums (right - DR, left - DL), currents on haulages (right - HR, left - HL), movement direction (MD), shearer speed, and location. Table 1 presents an example of collected data.

**Table 1:** Exemplary sensor data

| Timestamp | DR | DL | HR | HL | MD right | MD left | Location | Speed |
|---|---|---|---|---|---|---|---|---|
| | [A] | [A] | [A] | [A] | [0/1] | [0/1] | [m] | [m/s] |
| 24.01.10 03:43:17 | 67 | 61 | 21 | 17 | 1 | 0 | 33 | 4,30 |
| 24.01.10 03:43:18 | 65 | 60 | 19 | 20 | 1 | 0 | 33 | 4,30 |
| 24.01.10 03:43:19 | 69 | 61 | 20 | 20 | 1 | 0 | 33 | 4,30 |
| 24.01.10 03:43:20 | 65 | 62 | 20 | 20 | 1 | 0 | 33 | 4,30 |
| 24.01.10 03:43:21 | 74 | 61 | 20 | 20 | 1 | 0 | 33 | 4,30 |

One can notice that sensor data consists of mixed time-series data (binary and continuous) in a form unsuitable for process analysis with existing PM techniques. A cycle identifier is also absent in the raw data, making the PM analysis impossible.

In order to use sensor data for process mining, firstly, we have to preprocess our data to detect case ID (cycle ID) and subsequently transform raw sensor data into the higher-level activities (technological stages).

What is worth emphasising is that in our use case, cycle ID is not only important for PM but has crucial meaning for event abstraction. Without cycle ID, we cannot label low-level sensor data with rules describing technological stages. Therefore, we have a situation that is opposite to the most typical one, i.e., when we use set of events to correlate with known case ID. The mentioned circumstances make the task of cycle detection both difficult and interesting.

The first attempt to deal with cycle detection was the heuristic approach presented in [9]. Authors developed heuristics based on the longwall face length and distance range from the longwall's beginning and end. In most cases, the proposed heuristics correctly identified the cycle's start and end. However, in the case of severe data quality issues (e.g., changes in the location readings), the heuristic approach cannot detect cycles properly. That fact motivates us to develop an algorithm capable of dealing with the described issue.

## 3. Related work

Widely used methods related to event correlation issues can be classified into four categories [6,10]: (1) without any additional input, using only event names, (2) based on a process model, also using timestamps of events, (3) based on event similarity or case identifier in the log and (4) and relying on correlation conditions, using data attributes. We analysed them in the context of using raw sensor data instead of well-structured event logs with event names, timestamps, and other attributes.

In the first group, developed methods, e.g., presented in [11,12], assume the existence of event names in the event log. Thus, their usage is not possible with sensor data. Also, proposals described in [13,14] or [15,16] using process models to correlate events to sensor data are illusive if we do not have low-grained process model or adequate sensor-model mapping and unfortunately, this is the most frequent situation during work with sensor data.

Methods from the third category, assuming hidden case identifiers in the log, could be applicable to sensor data; however, their assumptions in real-life use cases very often are not fulfilled; besides, hidden case ID in data, activities names [17] or additional attributes [18] are not available. Also, Rule-Based Event Correlation using event similarity [19] also requires defined events and additional event attributes.

The last category of developed methods relies on correlation conditions using data attributes. Developed methods in this scope, i.e., [20–23] also assume an existence of higher level activities. Thus, their usage for sensor data is limited.

The contemporary methods for inferring missing entity identifiers, particularly concerning physical object movement within processes, employs Event Knowledge Graphs (EKG) [24]. Nevertheless, an underlying assumption within this approach is the prerequisite existence of activity names as input. Other approach for discrete event data in the maintenance domain, presented in [25] starts from single events and a set of events is clustered in different cases using heuristic created with domain experts and metrics enabling event labeling.

Upon reviewing the literature on event correlation within the process mining domain, it becomes evident that only a limited number of solutions facilitate case ID inference based on raw sensor data. This challenge is particularly pronounced within smart space domains such as homes, offices, and factories, as well as within activity recognition contexts [26]. In these domains, identifying individuals associated with activities and selecting relevant case notions can be challenging due to the absence of process-oriented segmentation in sensor logs. As outlined in the referenced review [26], the primary method employed to identify case IDs in smart space domains is time-based, utilising specific temporal markers within the data (e.g., midnight) or activities (e.g., "sleep") to partition the log into segmented traces. While this approach aligns well with research focused on Activities of Daily Living (ADL), it proves less suitable for industrial processes, which typically operate predominantly during daylight hours or across various working shifts. Another crucial disparity between ADL cases and industrial processes lies in the latter's utilisation of mixed data types (binary, continuous), rendering precise case ID identification a more demanding task.

The most present, consistent framework proposal for the domain-driven utilisation of sensor data in process mining is presented in [27]. The authors use an example of the beverage manufacturing process. For the case ID identification, authors propose looking for distinct characteristics in the machine data (like machine status, sequences of machine states or other attributes). The proposed algorithm uses activity groups for case ID inheritance. Thus, case ID identification requires activities to be known first, which is not available for many industrial processes (also for our mining use case).

To the best of our knowledge, only one example of case ID identification described in the literature does not require knowledge about process models or higher-level activities to label case ID in sensor data. The heuristic approach presented in [9] uses a selected variable – the location of the machine – in the form of a time series to detect unknown cycles in the cyclic process of the mining shearer operation.

The increasing popularity of PM in the industry and the lack of solutions for case ID identification require designing and implementation of new algorithms. Therefore, our goal is to develop an approach for case ID identification based on time-series patterns in sensor data, not requiring higher-level activities. In the next section, we describe the mining use case, which has motivated us to undertake research on this topic.

## 4. Our proposal

One of the possible approaches for cycle ID identification is to employ algorithms based on the analysis of short-term properties of analysed time series [28,29].

A pattern can be viewed as a sequence of segments (events and sample values) of different properties identified from analysed time series data. Catching dependencies between segments can be based on observations of typical (theoretical) patterns observed in real data, identified in an expert way. Based on such assumptions, we propose a rule-based algorithm dedicated to the detection of pre-defined patterns.

In the proposed approach, we assume that a pattern consists of the following sections:

- $S1$ – a local minimum (including neighbourhood) which can be viewed as the beginning of a sequence,
- $S2$ – a segment described by a significant change of the short-term average value of a series, of a fixed length;
- $S3$ – a short segment that represents a rapid change of a series value (including neighbourhood).

Thus, we mapped the above assumptions describing a pattern into an algorithm (see Algorithm 1).

To this aim, we used parameters for describing a shape of searched pattern(s) from analyzed series $Y$. $Y_{th}$ denotes a change in short-term statistical parameters of the series to be considered by the algorithm as a significant change, i.e. a part (component) of a pattern. $Lwz$ denotes current length of sections (length of a pattern) while $Lwz_{th}$ describes minimum length of sections, i.e. minimum number of samples considered as a pattern (small patterns are omitted). We also used $Ym_{th}$ describing the neighbourhood of a local minimum value of timeseries (close to 0). In a cycle ID detection, we assumed a use of a location variable describing the cyclic movement of the longwall shearer.

---

**Algorithm 1 Pattern (cycle) detection**

---

1: Read datataset (time series $Y$) of the length $L$ 2: Set/calculate the minimum length of a pattern as $Lwz_{th}$.

3: Set/calculate the minimum height of a pattern $Y_{th}$.

4: for $t$ = 1 to $L$ do

5: Calculate the short-term data sample mean value $Y_{msh}$

–    for the first pattern calculated within time interval started from 1 to $t$

–    for consecutive patterns - started for time $t_p$ after the end of detected previous pattern: from $t_p$ to $t$

6: IF $Lwz$ (a current length of sections) > $Lwz_{th}$ (a fixed minimum length of sections) AND

7: IF $Y_{msh}$ (a short-term mean value) > $Y_{th}$ (a threshold mean value) AND

8: IF a decrease in the mean value for further samples (time samples > $t$) is identified

9: is THAN a pattern is recognised for time $t$ (the last sample of this pattern is for time $t$) 10: end for

---

We identified two significant parameters for the described algorithm: $Y_{th}$ and $Lwz_{th}$. A number of calculations were carried out, including setting parameters related to the statistical properties of original input data and one-sample increments. To achieve high accuracy, we also tested parameter values fixed arbitrarily. In effect, for the original dataset, $Y_{th}$ was fixed to a value of 10.3 and $Lwz_{th}$ was fixed to a value of 100 in arbitrary/expert way.

Developed algorithm structure is simple; however presents relatively high accuracy in cyclic pattern detection. The following section describes the algorithm's operation on a real datasets.

## 5. Experiments and Results

Our mining dataset containing sensor readings without missing data (dataset called original data). In this case we used shearer location variable as input for cycle detection. In the original data, we observed high values of location (above 650m), which are obvious errors in data (in the longwall mining, the typical length of the longwall face does not exceed 300m). Thus, we removed the outliers, limiting the shearer location to 300m (dataset called as cleaned data).

We ran the developed algorithm (implemented in MATLAB environment) on the original and cleaned datasets. Results for original data are presented in Fig.2 while results for cleaned data are illustrated in Fig.3.

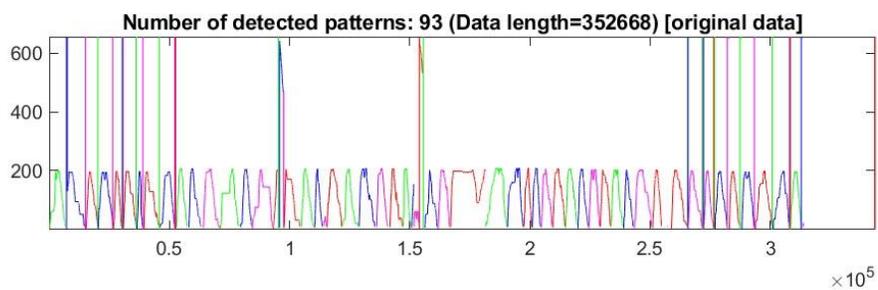

**Fig.2:** Results obtained for original data. (Remark: to distinguish subsequent patterns, different colours were used)

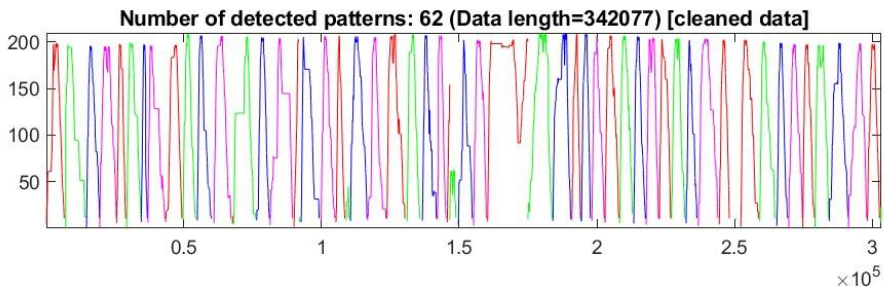

**Fig.3**: Results obtained for cleaned data.

As shown in Fig.2 and Fig.3, the algorithm can identify patterns in data. From the original data (containing 352668 samples of one second readings), we detected 93 patterns (shearers cycles), while in the cleaned data, the number of detected patterns is equal to 62. In Fig.2, we can observe that abnormal data values influence the efficiency of pattern detection. Untypical (in fact unreal) abrupt data changes affect short-term average values, thus a presence of additional patterns. It means that, on the one hand, the proposed algorithm is suitable for cycle ID detection; on the other hand, it is sensitive to abrupt changes in data (of high amplitude), which affects the change of short-term statistical properties of the data.

## 6. Evaluation

To evaluate the proposed algorithm, we conducted two fold experiments:

1. expert-based evaluation, based on labeled data provided by the domain expert,
2. domain-agnostic evaluation, based on dataset from manufacturing domain.

In both experiments for evaluation we used the F1 score [3].

First, we asked the domain expert to indicate the beginning and end of the cycle based on the process knowledge, taking into account separated labels for data outliers. In the analysed data set, expert indicated 56 shearer cycles and 33 labels for outlier data (89 cycles in total).

Subsequently, we compared discovered patterns with labels given by the domain expert and created a confusion matrix. The comparison visualisation in the form of a heatmap is presented in Fig.4.

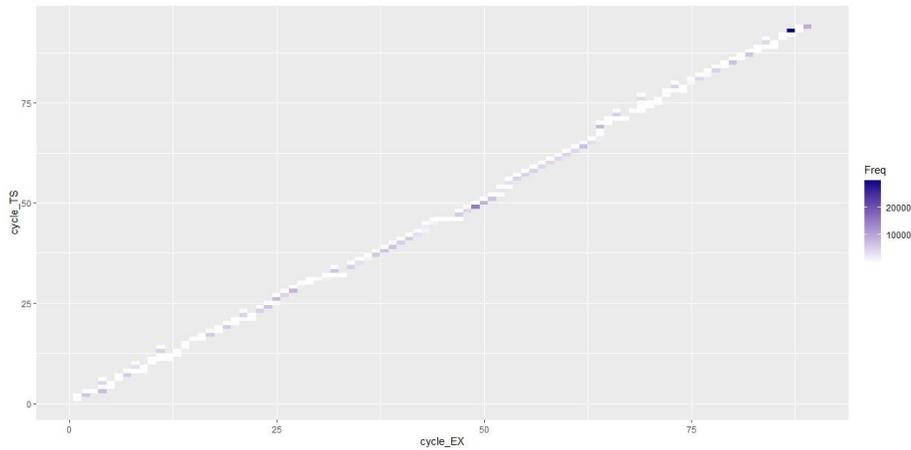

**Fig.4**: Comparison between expert labels (cycle_EX) and discovered patterns (cycle_TS) for original data

The heatmap shows the frequency of covering each pattern with expert labels. One can observe that discovered patterns cover expert labels quite well (with some exceptions related to abnormal data, which can be seen at the beginning (first 17 cycles), cycle 48, and cycles above 62).

In domain-agnostic evaluation, we used dataset from the CNC machine tools [1]. Originally, the mentioned dataset was used for energy prediction in the process [30]. The dataset comprises information about various tools (i.e., drills, mills, reamer) used to create a part (from aluminium alloy). We used this example because the process has a cyclic character denoted with a clear reading of energy consumption (given as time series - we focus on Power7), and data are labeled with tool ID (we chose SelectedTool), making a clear distinction of process instances. For CNC dataset, $Y_{th}$ was fixed to a value of 0.7 while $Lwz_{th}$ was fixed to a value of 100. The obtained results for cycle detection from CNC dataset are presented in Fig. 5.

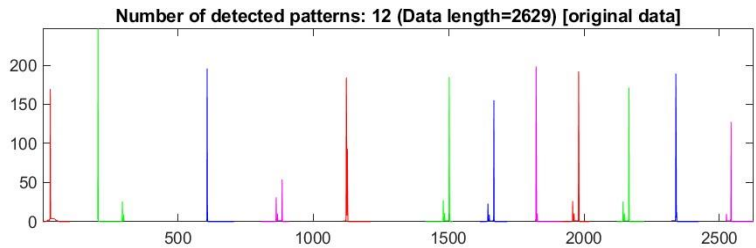

**Fig.5**: Results obtained for CNC dataset

In the next step, we calculated precision and recall, as proposed in [25], as well F1 score assuming as follows:

- *True Positive (TP)* – the number of samples in which the predicted CaseID match the distinct label indicated by domain expert (or selected variable);
- *False Positive (FP)* – the number of samples in which the predicted CaseID does not match the distinct label indicated by domain expert (or selected variable) because the samples belong to another case(s), denoting label mismatch;
- *False Negative (FN)* – the number of samples in which the CaseID predicted can not be matched to any label, denoting unnecessary labels.

Evaluation results are presented in Table 2.

**Table 2**: Evaluation metrics based on confusion matrix for datasets

| Dataset | Precision | Recall | F1 score |
|---|---|---|---|
| Shearer - Original | 97.7% | 96% | 96.8% |
| Shearer - Cleaned | 97.7% | 96.2% | 97% |
| CNC | 89.3% | 96.1% | 92.6% |

The presented results show the high accuracy of the developed algorithm in discovering patterns consistent with the expert labels. We repeated the evaluation procedure for cleaned data (without outlier values). Evaluation metrics for cleaned dataset are presented in Table 2. Not high differences between measures between the original and cleaned datasets result from a relatively small number of outliers indicated in the original dataset (10 591). In the comparison to the heuristic approach presented in [9] for original data set we obtained improvement of precision (2% percentage points), recall (17.7% pts) and F1 score (10.8% pts) and for cleaned dataset 2.2% pts, 17.7% pts and 11% pts respectively.

Evaluation results obtained for CNC dataset show that proposed approach can be also applied in caseID detection for cyclic processes in other domains.[2]

## 7. Conclusions and future work

In this paper, we present a novel algorithm for case ID identification that is useful for process mining based on sensor data without a registered case ID. The motivation for our work was lack of reliable method for case ID identification when only raw sensor data is available in the dataset.

Our algorithm is based on time-series analysis techniques to detect characteristic patterns in sensor data, enabling case ID detection. We ran our experiments on two datasets from a mining use case containing the location data of the longwall shearer. The first dataset contained errors (outliers) in sensor readings, and the second one contained filtered values of the location variable. Despite its simplicity, the algorithm effectively detected cycles in both original and cleaned data. We also applied our algorithm on CNC dataset, which confirmed its high efficiency on time-series data from other domain.

Although the developed algorithm efficiently recognises patterns in data, it is sensitive to abrupt changes in variable values. We analysed different sets of algorithm parameters related to the statistical properties of input time series and one-sample increments. However, valuable results were calculated for parameters fixed by an expert. So, the main parameters are determined arbitrarily. Therefore, further works will focus on an adaptive version of the algorithm, with parameters self-adjusting to the current properties of the analysed series (both statistical and frequency).

We are convinced that the presented algorithm is usable for other IoT use cases in which case ID is related to repetitive behaviour recorded in sensor data (e.g., cyclic movement of machines or personnel in a warehouse or production hall).

---

[2] Resources related to evaluation based on CNC dataset are provided here: https://dysk.agh.edu.pl/s/LPTsc3EYn8oGnyP